\documentclass[iop]{emulateapj}

\usepackage{epsfig}
\usepackage{longtable}
\usepackage{times} 
\newcommand{\sw}{$Swift$}

\newcommand \src {\mbox{IGR~J18219--1347}}

\newcommand \nh {N${\rm _H}$}

\shorttitle{Orbital period in IGR~J18219--1347}
\shortauthors{La Parola et al.}

\begin{document}

\title{\emph{Swift}-BAT hard X-ray sky monitoring unveils the orbital period  
of the HMXB  IGR~J18219--1347}

\author{V.\ La Parola\altaffilmark{1}, G.\ Cusumano\altaffilmark{1}, 
A.\ Segreto\altaffilmark{1}, A.\ D'A\`i\altaffilmark{2}, N.\ Masetti\altaffilmark{3}, V.\
D'Elia\altaffilmark{4,5}}

\altaffiltext{1}{INAF, Istituto di Astrofisica Spaziale e Fisica Cosmica,
        Via U.\ La Malfa 153, I-90146 Palermo, Italy}
\email{laparola@ifc.inaf.it}
\altaffiltext{2}{Dipartimento di Fisica e Chimica, Universit\`a di Palermo, via Archirafi 36, 90123,
Palermo, Italy}
\altaffiltext{3}{INAF - Istituto di Astrofisica Spaziale e Fisica Cosmica di Bologna, via Gobetti 101,
40129, Bologna, Italy}

 \altaffiltext{4}{INAF - Osservatorio Astronomico di Roma, Via Frascati 33, I-00040 Monteporzio Catone, Italy}
 \altaffiltext{5}{ASI-Science Data Centre, Via Galileo Galilei, I-00044 Frascati, Italy}

\label{firstpage}

\begin{abstract}
\src\ is a hard X-ray source discovered by INTEGRAL in 2010.
We have analyzed the X-ray emission of this source exploiting the BAT survey data up to March
2012 and the XRT data that include also an observing campaign performed in early 2012.
The source is detected at a significance level of $\sim 13$
standard deviations in the 88-month BAT survey data, and shows a strong variability along
the survey monitoring, going from high intensity to quiescent states. A timing analysis on 
the BAT data revealed an intensity modulation with a period of P$_0=72.44\pm0.3$ days. 
The significance of 
this modulation is about 7 standard deviations in Gaussian statistics. We interpret it as the 
orbital period of the binary system. The light curve folded at P$_0$ shows a sharp peak 
covering $\sim30\%$ of the period, superimposed to a flat level roughly consistent with zero.
In the soft X-rays the source is detected only in 5 out of 12 XRT observations, with the highest
recorded count rate corresponding to a phase close to the BAT folded light curve peak.
The long orbital period and the evidence that the source emits only during a small fraction
of the orbit suggests that the IGR~J18219-1347 binary system hosts a Be star.
The broad band XRT+BAT spectrum is well modeled with a flat absorbed power law with a high 
energy exponential cutoff at $\sim 11$ keV.


\end{abstract}



        \section{Introduction\label{intro}}

The INTEGRAL observatory \citep{Winkler03} with the IBIS/ISGRI telescope
\citep{Ubertini03,Lebrun03} and the Swift observatory \citep{Gehrels2004mn} with 
the Burst Alert Telescope (BAT, \citealp{Barthelmy2005:BAT}) are performing a
continuous monitoring of the sky in the hard X-ray energy band 
offering a long-term database for the activity of the X-ray sources and a large 
number of new detections of transient or very faint accreting sources. 
Most of the recently discovered high-mass X-ray binaries (HMXBs) are characterized by high 
local absorption ($\rm N_{H} > 10^{22}$ cm$^{-2}$) that prevented their 
detection by past soft X-ray monitoring. 
The BAT telescope is complementary to IBIS/ISGRI  for the study of the temporal behavior of 
these sources as it covers a fraction between 50\% and 80\% of the sky every day 
thanks to is large field of view (1.4 steradian half coded) and to its pointing
strategy.  The long (since 2004 December 15) and continuous monitoring has allowed to 
investigate the intrinsic emission variability and to search for the presence of 
long periodicities, unveiling the binary nature of many INTEGRAL sources 
(e.g. \citealp{corbet1, corbet2, 
corbet3, corbet4, corbet5, corbet6, corbet7, cusumano10, laparola10,  dai11}).

\begin{figure*}
\begin{center}
\centerline{\includegraphics[width=16.cm,angle=0]{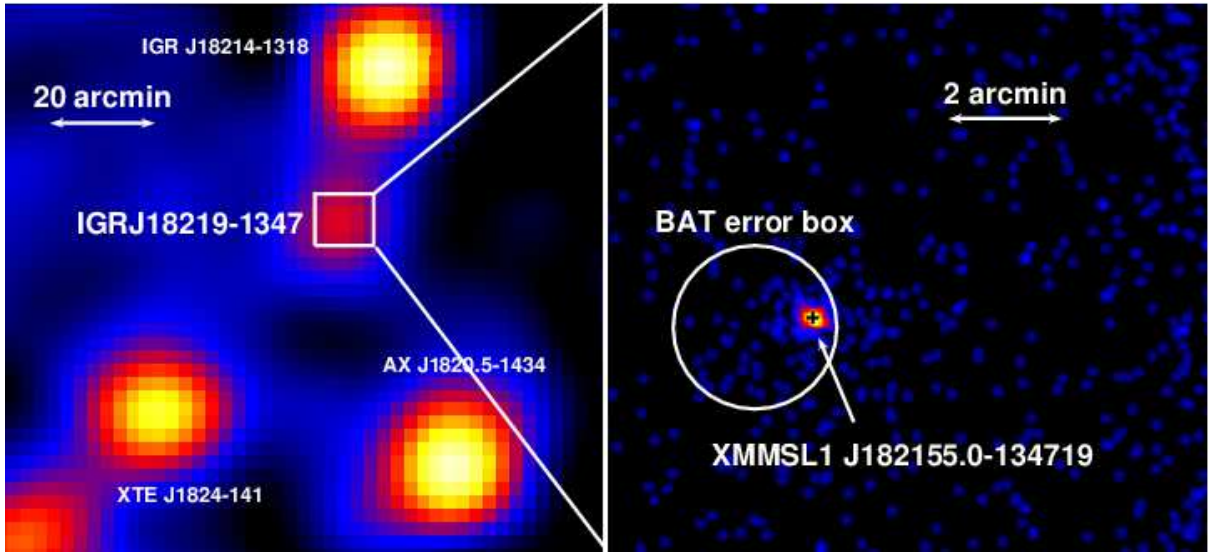}}
\caption[IGR~J18219--1347 sky maps]{ Left panel: 15--45 keV significance map in the 
region around IGR~J18219--1347.  
Right panel: 0.2-10 keV XRT image of observation 1 (see
table~\ref{log}).
                }
                \label{map} 
        \end{center}
        \end{figure*}

\begin{figure*}[!h]
\begin{center}
\centerline{\includegraphics[width=20cm]{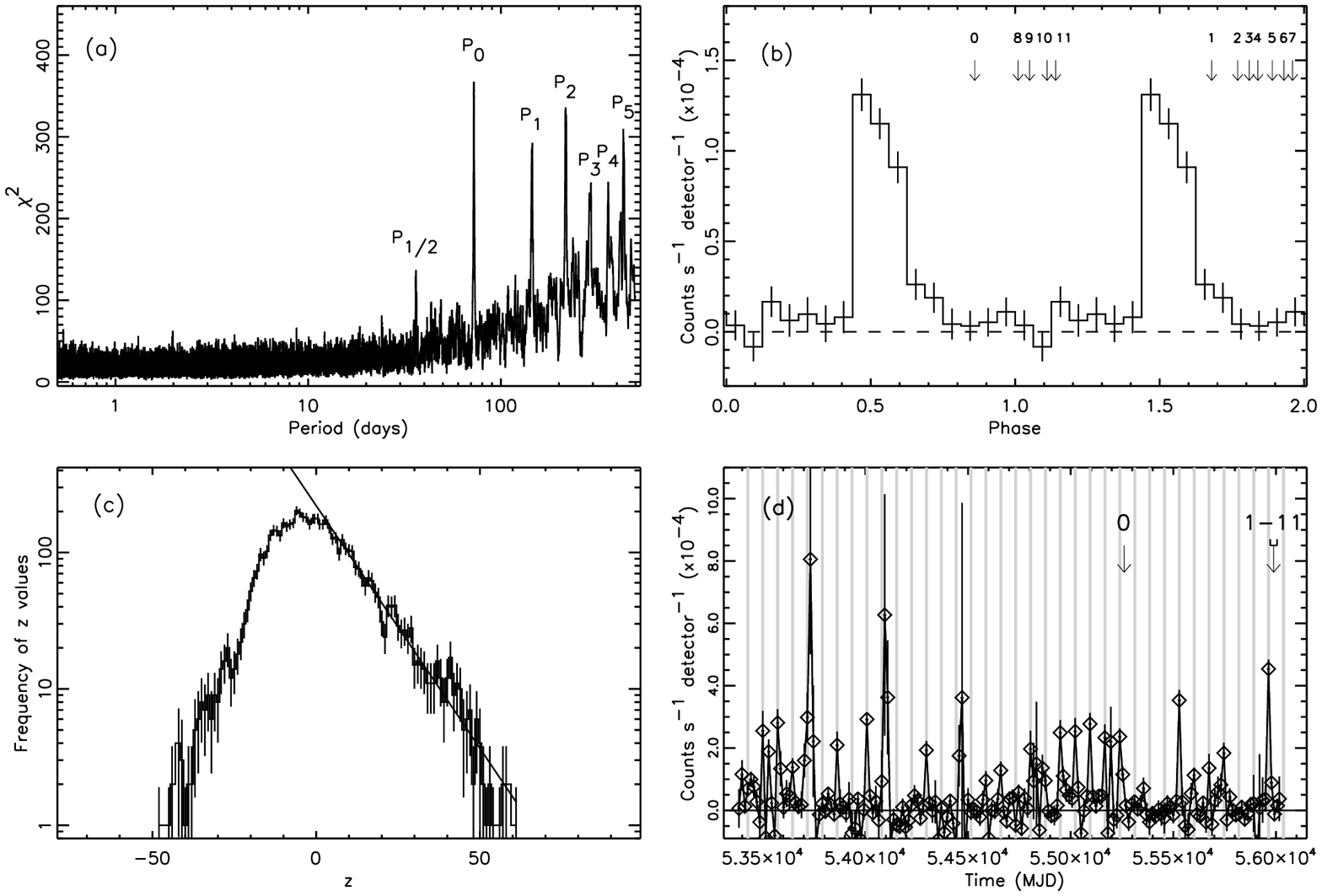}}

\vspace{0.5truecm}
\caption[]{{\bf a}: Periodogram of \sw/BAT (15--45\,keV) data for 
IGR~J18219-1347.
{\bf b}: Light curve folded at a period $P= 72.44$\,days, with 16 phase 
bins. 
Vertical arrows show the phase position corresponding to the epoch of each XRT
observation. 
{\bf c}: Distribution of z ($\chi^2$-$F_{\chi}$) 
selecting $\chi^2$ between 22 and 122 days  and excluding   
the values around the peak $P_0$.
The continuous line is the best fit obtained with an
exponential model  applied to the tail of the distribution (above z=22). 
{\bf d}: 15--45 keV BAT light
curve of IGR~J18219-1347 with a bin time of $P_0/5=14.49$ d. 
Vertical arrows show the epoch of the XRT observations (observations from 
Obs. 1 to 11 were performed within a time interval 33 days long). The shaded lines are
spaced by $P_0=72.44$ d and in phase with the peak position in Figure~\ref{period}b.
               }               
		\label{period} 
        \end{center}
        \end{figure*}

In this Letter we analyze the soft and hard X-ray data collected by Swift on 
IGR~J18219--1347. The source was discovered by INTEGRAL 
\citep{Krivonos} and it was associated with XMMSL1 J182155.0--134719 
because of a spatial coincidence. 
A 2MASS object (2MASS 18215463--1347232) found within the XMM 
counterpart position error box
was initially identified as the infrared counterpart of the IGR source \citep{Landi}. 
IGR~J18219--1347 was also detected by BAT \citep{Krimm} at the end of January 2012, 
with an intensity rising of a factor of 10 in
7 days: the average 17-60 keV flux was  $2\times 10^{-10}$ erg cm$^{-2}$ s$^{-1}$, 
more than 25 times the flux measured by INTEGRAL \citep{Krivonos}. 
The BAT detection triggered a Swift-XRT \citep{Burrows2005:XRTmn} follow up observing 
campaign allowing to obtain a refined position  with a localization
accuracy of 1.7'', confirming the soft X-ray counterpart XMMSL1 J182155.0--134719 and
rejecting the infrared counterpart candidate  2MASS 18215463--1347232 \citep{Krimm}. 
The \sw-XRT observations also revealed strong evidence for variability with an intensity
variation at least of a factor of 50.
A follow-up observation with Chandra found a soft X-ray counterpart at a 
position of RA$_{J2000}$ = 18h 21m 54.82s, Dec$_{J2000}$ = -13$^{\circ}$ 47' 26.7'' with a localization
accuracy of 0.64'', thus reducing the area of position uncertainty of a factor of 6
\citep{Karasev}. The Chandra source position was cross correlated with the
UKIDSS\footnote{http://www.ukidss.org/index.html} sky survey data allowing to 
identify an infrared source at RA$_{J2000}$ = 18h 21m 54.766s, 
Dec$_{J2000}$ = -13$^{\circ}$ 47' 26.77'', at a distance of $\sim$ 0.8'',
with magnitudes J=18.00, H=16.01, and K=14.44, as the likely counterpart. 
An accurate study of the profile of the image of this source showed,
however, that it is likely the superposition of two sources that cannot be separated because
of the spatial resolution of the infrared data.

Broad band (0.5--100 keV) spectral analysis combining \sw-XRT and INTEGRAL/IBIS data was modeled 
with an absorbed power law (\nh~ $ \rm \sim3\times10^{22} cm^{-2}$, photon 
index $\sim$1 and an exponential cutoff at $\sim$6 keV; \citet{Karasev}). 
The information derived from the broad band spectrum analysis and the strong variability 
suggested that IGR~J18219--1347 belongs to the class of the HMXBs.

This Letter is organized as follows. Section 2 describes the data reduction.
 Section 3 reports on the timing analysis. 
In Section 4 we describe the spectral analysis and
in Section 5 we briefly discuss our results. 

        \section{Data Reduction\label{data}}

The BAT survey data (15--150 keV) of the first 88 months of the \sw\ 
mission (2004 December -- 2012 March) were retrieved from the HEASARC public
archive\footnote{http://heasarc.gsfc.nasa.gov/docs/archive.html} and 
processed with a software \citep{segreto10} dedicated to the analysis of data of 
coded mask telescopes. The code performs screening, 
mosaicking and source detection and produces scientific products of any detected source.
IGR~J18219--1347 was detected in the 15--150 keV at a significance of 12.0 standard deviations, 
with a maximum of significance (12.7 standard deviations) in the 15--45 keV energy band.
Figure~\ref{map} (left) shows the 15--45 keV significance sky map (exposure time of 28.1 Ms) centered in
the direction of IGR~J18219--1347. 
We extracted the background subtracted spectrum of the source averaged over the entire survey 
and the light curve in the 15--45 keV energy range with the maximum available time 
resolution ($\sim 300$ s). The time tag of each bin of the light curve was corrected 
to the Solar system barycentre (SSB) using the JPL DE-200 ephemeris 
\citep{standish82} and the task 
{\sc EARTH2SUN}\footnote{http://heasarc.gsfc.nasa.gov/ftools/fhelp/earth2sun.txt}.

\begin{table*}
\scriptsize
\begin{tabular}{r l l l l l l l}
\hline
Obs \# & Instrument mode & Obs ID   & $T_{start}$ (TDB)    & $T_{elapsed}$  &Exposure (s) & Rate (c/s) & Orb.
Phase  \\   
        &     &          &  (MJD)               &  (s)           &  (s)  & &\\  \hline \hline
%
0 & XRT-PC           &    00031649001  & 55260.7956 & 17647.1   &1253.7 &$<$0.002& 0.86   \\
1 & XRT-PC       &    00032285001  & 55972.2877 & 39461.7   &1183.7 &0.17$\pm$0.01& 0.68 \\
2 & XRT-WT       &    00032285002  & 55978.6860 & 24777.1   &1786.2 &0.056$\pm$0.012 &0.77\\
3 & XRT-WT       &    00032285003  & 55981.7045 & 12850.7   &2691.8 &0.0682$\pm$0.009 &0.81\\
4 & XRT-WT       &    00032285004  & 55984.1737 & 24217.0   &2935.5 &$<$0.025 & 0.84\\
5 & XRT-WT       &    00032285005  & 55987.6437 & 13019.6   &1476.7 &0.027$\pm$0.009 &0.89\\
6 & XRT-WT       &    00032285006  & 55990.5850 & 24750.7   &1247.3 &0.022$\pm$0.011 & 0.93\\
7 & XRT-WT       &    00032285007  & 55993.0593 &  7396.7   &2797.1 &$<$0.022 & 0.96\\
8 & XRT-WT       &    00032285008  & 55996.0047 &  5457.5   &1462.8 &$<$0.027 & 0.01\\
9 & XRT-WT       &    00032285009  & 55999.0754 & 24613.9   &2604.3 &$<$0.021 &0.05\\
10 & XRT-WT       &    00032285010  & 56003.765 & 18656.8   &3048.0 &$<$0.021 &0.11\\
11 & XRT-WT       &    00032285011  & 56005.7693 &  7452.3   &2814.0 & $<$0.024&0.14\\ \hline
\end{tabular}
\caption{Log of the Swift-XRT observations. \label{log}}
\end{table*}

\sw-XRT observed the field of IGR~J18219--1347  on 2010 March 5
 in Photon Counting (PC) mode  without detecting any source (Obs \# 0 in table~\ref{log}).
Two years later (2012 February 15), following the detection of the source as a transient by BAT, a one month 
target of opportunity campaign was activated. The source was observed 
in  PC mode in the first observation (obs \# 1 in table~\ref{log}) and in Windowed Timing (WT) mode 
\citep{hill04} in the following observations (obs \# 2 to 11 in table~\ref{log}). 
The XRT data were processed with standard procedures
({\sc xrtpipeline v.0.12.4}), filtering and screening criteria, using
ftools in the Heasoft package (v 6.8). 
We adopted standard grade filtering 0-12 and 0-2 for PC and WT data, respectively.
IGR~J18219--1347 was detected only in 5  out of 11 observations.
Table 1 reports the details on the \sw-XRT observations and the relevant count rate. 
Figure~\ref{map} (right) shows the 0.2--10 keV XRT image  
from ObsID 00032285001 where a source consistent in position to the likely soft X-ray 
counterpart  XMMSL1 J182155.0--134719
is detected with a significance of $\sim 14$ standard deviations. No other sources are
detected inside the BAT error box.

The source events for data collected in PC mode (Obs \# 1) 
were extracted from a circular region of 20 pixel radius (1
pixel = 2.36'') centered on the source position as determined with
{\sc xrtcentroid} while, for data  collected in WT mode, events were extracted 
only from obs. 2, 3, 5 and 6 (see Table 1) by selecting a 20 pixel 
wide portion of the WT strip centered on the source position.
The background for spectral analysis was extracted from an annular region
centered on the source with radii of 70 and 130 pixels for the observation in PC mode 
and from two strip regions (symmetrical with respect to the source position) sufficiently
offset ($>2$ arcminutes) from the source to avoid any contamination
from its PSF wings in WT mode. 
We used the official BAT spectral redistribution
matrix\footnote{http://heasarc.gsfc.nasa.gov/docs/heasarc/caldb/data/swift/bat/index.html}
and the XRT spectral redistribution matrix v013.
XRT ancillary response files were generated with 
{\sc xrtmkarf}\footnote{http://heasarc.gsfc.nasa.gov/ftools/caldb/help/xrtmkarf.html} 
 Spectral fits were performed using {\sc xspec} v.12.
The source and background spectra of the observations in WT mode
were combined to obtain a single spectrum, and the ancillary files were combined using {\sc addarf}
weighting them by the exposure times of the relevant spectra. Finally, spectra
were rebinned with a minimum of 20 counts per energy channel, in order to allow the use of
the $\chi^2$ statistics. 

        \section{Timing analysis\label{sfxt7:timing}}

We analyzed the long term BAT light curve to search for periodic intensity modulation 
{using the epoch folding method \citep{leahy83}}.
The 15--45 keV BAT light curve was folded with different trial periods P from 0.5\,d to 500\,d
with a step of $P^{2}/(N \,\Delta T)$, where $N=16$ is the number of trial profile 
phase bins and $\Delta T=$211.7  Ms is the data time span. To build profiles for each
trial period we applied a weighing procedure (e.g. \citealp{cusumano10}) 
suitable for background dominated data with a large span of count rate errors.
Figure~\ref{period}a shows the periodogram, where several features emerge. 
The highest one has a $\chi^2$ value of $\sim358$ and corresponds to a period of 
P$_0=72.44\pm0.3$ d. From fitting the peak profile with a Gaussian 
function, we derived a P$_0$ centroid and a standard deviation
of 72.44 d and 0.3 d, respectively.
We see also other evident features at periods multiples of P$_0$ (P$_1$, P$_2$, P$_3$, 
P$_4$, P$_5$ in Fig~\ref{period}a) and at the sub-multiple  $P_0/2$. 
The intensity profile (Fig~\ref{period}b) 
folded at P$_0$ with T$_{\rm epoch}$=54619.3125 shows a peak
that covers 30\% of the period, over a flat level consistent with zero.
To evaluate the phase position of the peak centroid we have built a pulse profile 
folding BAT data with  P$_0$, $N=30$ phase bins and fit the peak  
with a Gaussian model: the centroid is at phase $0.510\pm 0.004$ corresponding to 
MJD $(54656.26\pm0.29)\pm nP_0$. The presence of the feature at $P_0/2$
is a direct consequence of the sharp and narrow peak shown in Figure~\ref{period}b: 
folding with a period equal to P$_0/2$ this peak will be added coherently to the profile 
every two cycles, producing a feature in the periodogram with an intensity significantly
lower with respect to the main feature.

As a consequence of the source variability and of the presence of a periodic 
signal, the average $\chi^2$ in 
the periodogram is far from the average value expected for white noise $(N-1)$ 
and the $\chi^2$ statistics cannot  be applied.
The significance of the observed feature shall be evaluated with respect to
the average level of the periodogram noise. For this reason, we fitted
the periodogram with a second order polynomial and subtracted the best fit 
($F_{\chi}$) to the $\chi^2$ distribution. The $z=\chi^2-F_{\chi}$ 
distribution has a value of $\sim 304$ at $P_0$. We therefore 
have built the histogram of the $z$ distribution (Figure~\ref{period}c) 
extracting the values only from 22  to 122~d (where the noise 
level is quite consistent with the noise level at $P_0$)
and excluding the interval around $P_0$ and around $P_0/2$.
We fit the tail ($z >10$) of this distribution  with an exponential 
function. The resulting best fit model is plotted in  Figure~\ref{period}c. 
We have evaluated the area under the histogram dividing it into two parts: from 
its left boundary up to $z=10$ we have summed the contribution of each 
single bin; beyond $z=10$ up to infinity we have integrated the best fit 
exponential model. Therefore, we evaluated the integral of the best-fit exponential 
function beyond 304 and normalized it to the total area of the histogram. 
The result ($6.7\times10^{-12}$) is the probability of chance occurrence of a 
$z$ equal to or larger than 304 or a $\chi^2$ equal to or larger than 358 and it corresponds to a 
significance for the detected feature of $\sim 7$ standard deviations in Gaussian statistics.

Figure~\ref{period}d shows the 15--45 keV light curve of IGR~J18219-1347 with a 
bin time of P$_0/5$. 
To visually show the periodicity in the BAT light curve, 
we over-plotted vertical shaded bars spaced by P$_0$  and in phase with the peak
of the BAT folded profile (Fig.~\ref{period}b).
Table~\ref{log} reports the phase corresponding to each XRT observation relevant to the BAT light curve
profile in Figure~\ref{period}b.
We observe that the pointing with the highest count rate corresponds to the phase closest to the
peak, while upper limits correspond to orbital phases farther from the peak.
The statistics of the source in the XRT dataset both in PC and in WT mode is too low to allow for
timing analysis finalized to the search for a pulsation.

\section{Broad band spectral analysis\label{spetra}}

Broad band spectral analysis was performed using the XRT data relevant to the observations
with the source detected above 2 standard deviations (Obs.~1 in PC mode and Obs.~2, 3, 5
and 6 in WT mode that were summed into a single spectrum because the statistics of each single 
WT dataset is too low to build a significant spectrum; see table~\ref{log} and Sect.~2) and
the BAT hard X-ray spectrum  averaged over  the 88 months of monitoring, selecting only intervals
corresponding to the phase of the peak. We assumed no
significant spectral variability among the XRT observations and during the BAT monitoring.
A preliminary analysis was made to verify that this assumption was indeed  valid.
The XRT spectra in PC mode  and  WT mode were fitted simultaneously with an absorbed power-law 
with the absorption hydrogen column  and the photon index 
parameters forced to have the same value in
both datasets. The best fit residuals showed the same trend for both datasets, with a
 best fit photon index equal to $-0.2_{-0.4}^{+0.9}$ and absorbing column density equal 
 to $3.3_{-2}^{+3} \times 10^{22} cm^{-2}$. 
Six BAT spectra were produced to verify the assumption of non-variability both during the
88-month monitoring (dividing it into four 22-month long intervals) and/or during different phase
intervals of Figure~\ref{period}b (selecting the two spectra corresponding to phase intervals 
0.4375--0.625 and 0.625--1.375). These spectra were fitted with a power law 
model and, as above, the photon index was constrained to have the same value  for all the 
spectra without any constraint for the normalization parameters.  The best fit residuals show 
the same trend for all the datasets, with best fit photon index $2.64\pm0.14$.

\begin{figure}
\begin{center}
\centerline{\includegraphics[width=6cm,angle=-90]{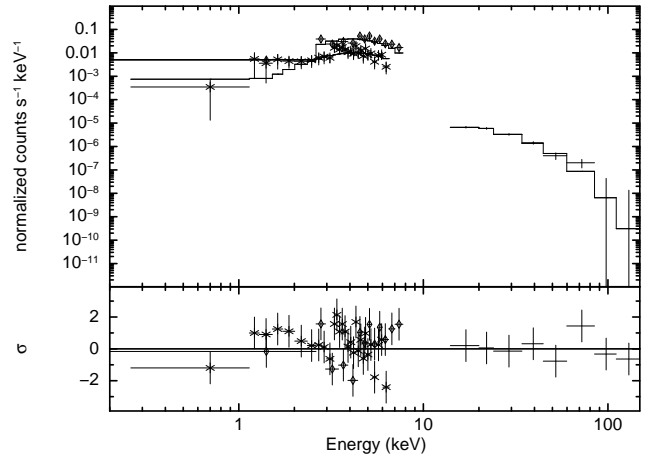}}
\caption[IGR~J18219-1347 XRT and BAT spectra]{{\bf Top panel}:
IGR~J18219-1347 
XRT (PC: diamond points; WT: star points) and BAT spectra and best fit model. 
{\bf Bottom panel}: Residuals in unit of standard deviations.
}
                \label{spec} 
        \end{center}
        \end{figure}

Therefore, assuming no significant spectral variability during the BAT monitoring and between the  
XRT spectra we performed a broad band spectral analysis coupling the soft X-ray spectra with the BAT 
spectrum selected in phase with the peak in the folded profile (phase intervals 0.4375 to 
0.625 in Figure~\ref{period}b) introducing in the model a multiplicative factor to 
take into account both an 
intercalibration factor between the two telescopes and the different average flux level among the three 
datasets.  An absorbed power law model gave  an 
unacceptable $\chi^2$ of 89.21 (52 d.o.f.). This 
was expected because of the difference in photon index found in the analysis performed separately 
for the XRT and BAT datasets. The broad-band spectrum resulted indeed well fitted 
($\chi^2$=61.7 (51 d.o.f.) adding to the power law a high 
energy exponential cutoff ({\tt phabs*cutoffpl} model).
Figure~\ref{spec} shows the combined XRT and BAT energy distribution 
with best fit model (top panel) and residuals in units of standard deviations (bottom panel).
Table~\ref{fit} reports the best fit parameters (quoted errors are given at
90\% confidence level).


%
\section{Discussion and conclusions\label{discuss}}

We have presented the results obtained from the analysis of the data collected 
by \emph{Swift} on IGR~J18219-1347. The first 88 months of BAT survey data 
show a strong emission variability in the hard X-ray energies with the source going from  
quiescent to high emission states. The timing analysis performed on the BAT data 
unveils a periodic modulation of P$_0$=72.44 days that we interpret as the orbital period of
binary system. 
The profile obtained folding the BAT light curve at P$_0$ shows a flat plateau consistent with the
source being in a quiescent state, and a  narrow peak lasting only 30\% of the orbital period.
The brightness enhancement on 2012 Feb. 6 (MJD 55963), observed by BAT and reported by 
\citet{Krimm}, correspond to phase 0.54, fully consistent with the phase of the peak in 
the folded light curve. 
The one-month XRT observing campaign that followed this detection shows that the soft X-ray 
emission is in agreement with the 
hard X-ray modulation, with the source {\bf strongly} detected only in the observations close 
to the BAT folded light curve peak {\bf (obs 1 to 3)} and with no or marginal detections 
(Obs 5 and 6) in the observations corresponding to the plateau. {\bf The non-detection in Obs 4 is}
due to a higher background level (a factor of 2 higher than in Obs 5 and 6).

The long orbital period and the evidence that the source emits only during a small fraction of the
orbit suggests that the high mass companion of IGR~J18219-1347 is a Be star. Be/X-ray binaries
are indeed the most numerous class of HMXB and are characterized by wide orbits with moderate
eccentricities, as opposite to supergiant X-ray binaries that show circular orbits with 
periodicities shorter than $\sim10$ days (e.g. \citealp{Liu06}). 
In a Be/X-ray binary system the accretion onto the neutron star is driven by mass capture from an 
equatorial disk around the equatorial plane of the Be star \citep{okazaki01}. This is favored at 
the periastron passage and causes the so-called type-I outbursts, while for most of the orbit a large
orbital distance prevents the accretion, keeping the source in a quiescent state. 
This is consistent with the overall behaviour observed in the BAT light curve of IGR~J18219-1347
(Fig.~\ref{period}d), with the source emission characterized by periodical emission peaks. As observed
for Be/Xray binary systems \citep{reig99, reig11}, the intensity
of these peaks show a large variability and is related to the amount of disk material captured for 
accretion during the periastron passage.
\citet{corbet86} showed that in  Be/X-ray binaries the orbital period tends to be correlated with 
the spin period. If IGR~J18219-1347 is indeed a member of this class, we can expect a spin period
longer than 10 s.

The broad-band (0.2--150 keV) spectrum of IGR~J18219--1347 
is well modeled with a flat absorbed  power law with  a high
energy exponential cutoff at $\sim 11$ keV. The column density
is $\rm \sim 4.3 \times 10^{22}~cm^{-2}$, which is a factor of $\sim 3$ larger than the
Galactic value in the direction of the source ($\rm 1.3 \times 10^{22} cm^{-2}$
\citealp{dickey90}). This suggests an additional intrinsic absorption in the environment
of the binary system. These spectral results are in agreement with those reported 
by \citet{Karasev} from the analysis of the XRT/Swift and IBIS/INTEGRAL data.

\begin{table}
\caption{Best fit spectral parameters.  \label{fit}}
\begin{tabular}{ r l l}
\tableline
Parameter & Best fit value & Units    \\ \tableline \tableline
N$_{\textrm{H}}$& $4.3^{+3.8}_{-1.7} \times 10^{22}$ & $\rm cm^{-2}$\\
$\Gamma$  &$-0.1^{+1.1}_{-0.6}$&          \\
$E_{cut}$ &$11^{+8}_{-2}$ & keV\\
$N$       &$ (7^{+9}_{-5})\times 10^{-4}$ &ph $\rm keV^{-1} cm^{-2} s^{-1}$ at 1 keV \\
$\rm C_{XRT-WT}$&$0.26\pm0.05$&\\
$\rm C_{BAT}$&$0.33^{+0.48}_{-0.14}$&\\
$\rm F$ (0.2--10 keV)&$4.25\times 10^{-11}$& erg s$^{-1}$ cm$^{-2}$\\
$\rm F$ (15--150 keV)&$4.23\times 10^{-11}$& erg s$^{-1}$ cm$^{-2}$\\
$\chi^2$   &61.7 (51 d.o.f.) & \\ \tableline
\end{tabular}
\tablecomments{$\rm C_{XRT-WT}$ and $\rm C_{BAT}$ are the
constant factors to be multiplied to the model in order to match the XRT-WT and BAT data,
respectively. We report unabsorbed fluxes for the standard XRT (0.2--10 keV) and BAT (15--150 keV)
energy bands.}

\end{table}

\acknowledgments

This work was supported in Italy by ASI grant I/011/07/0. 


\label{lastpage}

\end{document}